\documentclass[%
reprint,
superscriptaddress,
amsmath,amssymb,
aps,
prb,
longbibliography,
]{revtex4-1}

\usepackage{pdfpages} 
\usepackage{dirtytalk}
\usepackage{mathtools}
\usepackage{pdfpages}
\usepackage{mhchem}
\usepackage{graphicx}
\usepackage{dcolumn}
\usepackage{bm}

\usepackage{hyperref}
\hypersetup{
colorlinks = true,   
urlcolor = blue,    
linkcolor = magenta,   
citecolor = blue  
}
\usepackage{float}
\usepackage{color}

\makeatletter
\AtBeginDocument{\let\LS@rot\@undefined}
\makeatother

\begin{document}
 \title{Surface Bogoliubov-Dirac cones and helical Majorana hinge modes\\ in superconducting Dirac semimetals}
 \author{Majid Kheirkhah}
 \affiliation{Department of Physics, University of Alberta, Edmonton, Alberta T6G 2E1, Canada}
 \author{Zheng-Yang Zhuang}
  \affiliation{School of Physics, Sun Yat-Sen University, Guangzhou 510275, China}
 \author{Joseph Maciejko}
 \affiliation{Department of Physics, University of Alberta, Edmonton, Alberta T6G 2E1, Canada}
 \affiliation{Theoretical Physics Institute, University of Alberta, Edmonton, Alberta T6G 2E1, Canada}
 \author{Zhongbo Yan}
 \email{yanzhb5@mail.sysu.edu.cn}
 \affiliation{School of Physics, Sun Yat-Sen University, Guangzhou 510275, China}
\date{\today}
\begin{abstract}  
In the presence of certain symmetries, three-dimensional Dirac semimetals can harbor not only surface Fermi 
arcs, but also surface Dirac cones. Motivated by the experimental observation 
of rotation-symmetry-protected Dirac semimetal states in iron-based superconductors, we investigate the potential 
intrinsic topological phases in a $C_{4z}$-rotational invariant superconducting Dirac semimetal with 
$s_{\pm}$-wave pairing. When 
the normal state harbors only surface Fermi arcs on the side surfaces, we 
find that an interesting gapped superconducting state with a quartet of Bogoliubov-Dirac  cones on each side surface
can be realized, even though the first-order topology of its bulk is trivial.
When the normal state simultaneously harbors surface Fermi arcs and surface Dirac cones, 
we find that a second-order time-reversal invariant topological superconductor with helical Majorana 
hinge states can be realized. The criteria for 
these two distinct topological phases have a simple geometric interpretation in terms 
of three characteristic surfaces in momentum space. 
By reducing the bulk material to a thin film normal to the axis of rotation symmetry, we further find that a two-dimensional first-order time-reversal invariant topological superconductor can be realized 
if the inversion symmetry is broken by applying a gate voltage. Our work reveals that diverse topological 
superconducting phases and types of Majorana modes can be realized in superconducting Dirac semimetals.
\end{abstract}

\pacs{Valid PACS appear here}
\maketitle
\section{Introduction}
Topological superconductors (TSCs) are a class of novel phases with exotic gapless boundary excitations known as 
Majorana modes~\cite{read2000,kitaev2001unpaired}. Over the past decade, the pursuit of TSCs and 
Majorana modes in real materials has attracted a great amount of
enthusiasm~\cite{alicea2012new,leijnse2012introduction,Beenakker2013,stanescu2013majorana,Elliott2015,sato2016majorana,aguado2017majorana,Haim2019,Jack2021review}, owing to their exotic properties and potential applications in topological quantum computation~\cite{ivanov2001,Alicea2011non,nayak2008review,sarma2015majorana}. 
Very recently, an important form of progress on the theoretical side is 
the birth of the concept named higher-order TSCs~\cite{Benalcazar2017,Schindler2018,Benalcazar2017prb,Song2017,Langbehn2017hosc,Khalaf2018,Geier2018,Zhu2018hosc,Yan2018hosc,Wang2018hosc,Wangyuxuan2018hosc,Yan2019hosca}, which not only 
enriches the physics of TSCs, but also provides new perspectives for the realization and
applications of Majorana modes~\cite{Shapourian2018SOTSC,Hsu2018hosc,Liu2018hosc,Wuzhigang2019hosc,Zhang2019hinge,Zhang2019hoscb,Volpez2019SOTSC,Zhu2019mixed,Peng2019hinge,Ghorashi2019,Yan2019hoscb,Bultinck2019,Franca2019SOTSC,Pan2019SOTSC,Majid2020hosca,Majid2020hoscb,Wu2020SOTSC,Hsu2019HOSC,Wu2020BOTSC,Laubscher2020hosc,Tiwari2020,Ahn2020hosc,Bitan2020hosc,Li2021BTSC,Niu2020hosc,Wuxianxin2021hosc,Fu2021hinge,Luo2021hosc,Jahin2021,Qin2021SOTSC,Tan2021hosc,You2019,Zhang2020mzm,Zhang2020tqc,Pahomi2020braiding,Bomantara2020tqc,Lapa2021,PhysRevResearch.3.023007,roy2021mixed}. The most prominent difference between 
conventional TSCs and their higher-order counterparts lies in the bulk-boundary 
correspondence, or more precisely, the codimension $d_{c}$ of the gapless Majorana modes at the boundary. 
To be specific, a conventional TSC has $d_{c}=1$, while an $n$th-order
TSC has $d_{c}=n\geq2$. Conventional TSCs are thus also dubbed as
first-order TSCs. 
One direct significance of this extension is that  a lot of systems previously thought to be trivial 
in the framework of first-order topology are recognized to be nontrivial  
in the framework of higher-order topology.

Because of the scarcity of odd-parity superconductors in nature, the realization of 
both first-order and second-order TSCs heavily relies on  materials with strong spin-orbit coupling or topological 
band structure \cite{fu2008,lutchyn2011, oreg2010helical,sau2010,alicea2010,Yan2018hosc,Wang2018hosc,Zhu2019mixed,Majid2020hoscb}. By far, most experiments in this field 
have focused on the realization of first-order TSCs in various kinds of
heterostructures which simultaneously consist of three ingredients, namely, spin-orbit coupling,
magnetism or external magnetic fields,  and $s$-wave superconductivity~\cite{Mourik2012MZM,rokhinson2012fractional,das2012zero,deng2012anomalous,finck2013,nadj2014observation,Sun2016Majorana,Deng2016Majorana,Fornieri2019,Ren2019}.
Despite steady progress in experiments, the complexity of such heterostructures
and the concomitant strong inhomogeneity
make a definitive confirmation of the expected Majorana modes remain elusive~\cite{Zhang2019perspective,Frolov2020,Yu2021}.
Since these common shortcomings of heterostructures shadow the
pursuit of Majorana modes and are quite challenging to overcome in the short term,
intrinsic TSCs become highly desired to make further
breakthroughs. Remarkably, the band structures of a series of iron-based superconductors 
have recently been observed to host both topological insulator states  and rotation-symmetry-protected Dirac semimetal (DSM) states 
near the Fermi level \cite{zhang2018iron, zhang2019multiple}.
Since the coexistence of topological insulator states and superconductivity provides 
a realization of the Fu-Kane proposal~\cite{fu2008} in a single material, the potential existence of 
Majorana zero modes in the vortices of these iron-based superconductors has attracted 
great attention~\cite{wang2018evidence,Liu2018MZM,machida2019zero,kong2019half,Zhu2019MZM,Jiang2019vortex,Qin2019vortex2,Chiu2020vortex,Ghazaryan2020vortex,wu2021vortex,Majid2021vortex}. In addition, it turns out that the combination of
topological insulator states and unconventional $s_{\pm}$-wave pairing also makes 
these iron-based superconductors promising for the realization of intrinsic 
higher-order TSCs~\cite{Zhang2019hinge,Majid2021vortex}. Compared with the topological insulator states, we notice 
that the  DSM states in these iron-based superconductors 
have been explored much less~\cite{Qin2019vortex,Konig2019votex,Yan2020vortex,Kawakami2019}.

Motivated by the above observation, we explore the potential intrinsic
topological phases in superconducting DSMs with $s_{\pm}$-wave pairing.
However, instead of considering a realistic but complicated 
Hamiltonian to accurately produce the band structure of one specific 
iron-based superconductor, we will take a minimal-Hamiltonian approach
for generality, so that the results can be applied 
to all DSMs with the same symmetry and topological properties. To be relevant to iron-based superconductors, in this paper we focus on DSMs protected by $C_{4z}$-rotation symmetry \cite{Yang2014classification}.
For DSMs, while the low-energy physics in the bulk can be universally described by linear 
continuum  Dirac Hamiltonians, the gapless states on the boundary, however, 
are sensitive to the details of the full lattice Hamiltonian. An important fact is 
that both surface Fermi arcs and surface Dirac cones are symmetry allowed in 
DSMs \cite{Kobayashi2015Dirac,Kargarian2016,Yan2020vortex,Wieder2020}. As a 
consequence, we find that depending on whether surface Fermi arcs and surface Dirac cones coexist or not, 
a second-order time-reversal invariant TSC with helical Majorana hinge modes
or an interesting gapped phase with a quartet of Bogoliubov-Dirac  cones on each side surface
can be realized in the superconducting DSM, respectively. By reducing the dimension of the second-order time-reversal invariant TSC 
from three dimensions to a thin film, we 
find that a first-order time-reversal invariant TSC can be realized if the inversion symmetry is broken by applying an external 
gate voltage. These findings suggest that the 
superconducting DSM on its own can realize a diversity of intrinsic TSCs. 

The paper is organized as follows. In Sec.~II, the topological 
properties of the normal state are investigated. 
In Sec.~III, we show that two superconducting phases with distinct topological boundary states 
can be realized in superconducting DSMs. In Sec.~IV, 
we show that thin films 
of a superconducting DSM can realize first-order time-reversal invariant TSCs. Finally, we conclude with a discussion in Sec.~V. Some calculation details are relegated
to Appendices A and B. 

\section{Topological properties of the normal state}
We start with the DSM Hamiltonian which, in the basis
$\psi^{\dag}_{\bm{k}}=(c_{a,\uparrow,\bm{k}}^{\dag}
,c_{b,\uparrow,\bm{k}}^{\dag},c_{a,\downarrow,\bm{k}}^{\dag},c_{b,\downarrow,\bm{k}}^{\dag})$, reads \cite{Yang2014classification} 
\begin{eqnarray}
H_{\rm DSM}(\bm{k})&=&[m-t(\cos k_{x}+\cos k_{y})-t_{z}\cos k_{z}]\sigma_{z}\nonumber\\
&&+\lambda\sin k_{x}s_{z}\sigma_{x}+\eta \sin k_{z}(\cos k_{x}-\cos k_{y})s_{x}\sigma_{x}\nonumber\\
&&-\lambda\sin k_{y}\sigma_{y}+2\eta\sin k_{x}\sin k_{y}\sin k_{z}s_{y}\sigma_{x}, \label{Hamiltonian}
\end{eqnarray}
where the Pauli matrices $\sigma_{i}$ and $s_{i}$ act on the orbital ($a,b$) and 
spin ($\uparrow, \downarrow$) degrees of freedom, respectively. For notational simplicity,
the lattice constants are set to unity throughout this paper, and identity
matrices are always made implicit. 
The Hamiltonian simultaneously has  time-reversal symmetry ($\mathcal{T}=is_{y}\mathcal{K}$,
with $\mathcal{K}$ denoting complex conjugation), inversion symmetry ($I=\sigma_{z}$), and 
$C_{4z}$-rotation symmetry ($C_{4z}=\text{diag}\{e^{-3i\pi/4},e^{-i\pi/4},e^{3i\pi/4},e^{i\pi/4}\}$), which thus allows the presence of robust Dirac points 
on the rotation symmetry axis. It is easy to find that Dirac points will appear as long as 
the band inversion surface (BIS), which is defined as the zero-value contour of  $m-t(\cos k_{x}+\cos k_{y})-t_{z}\cos k_{z}$
in momentum space,  encloses one time-reversal invariant momentum.

\begin{figure}[t!]
\centering
\includegraphics[scale=0.266]{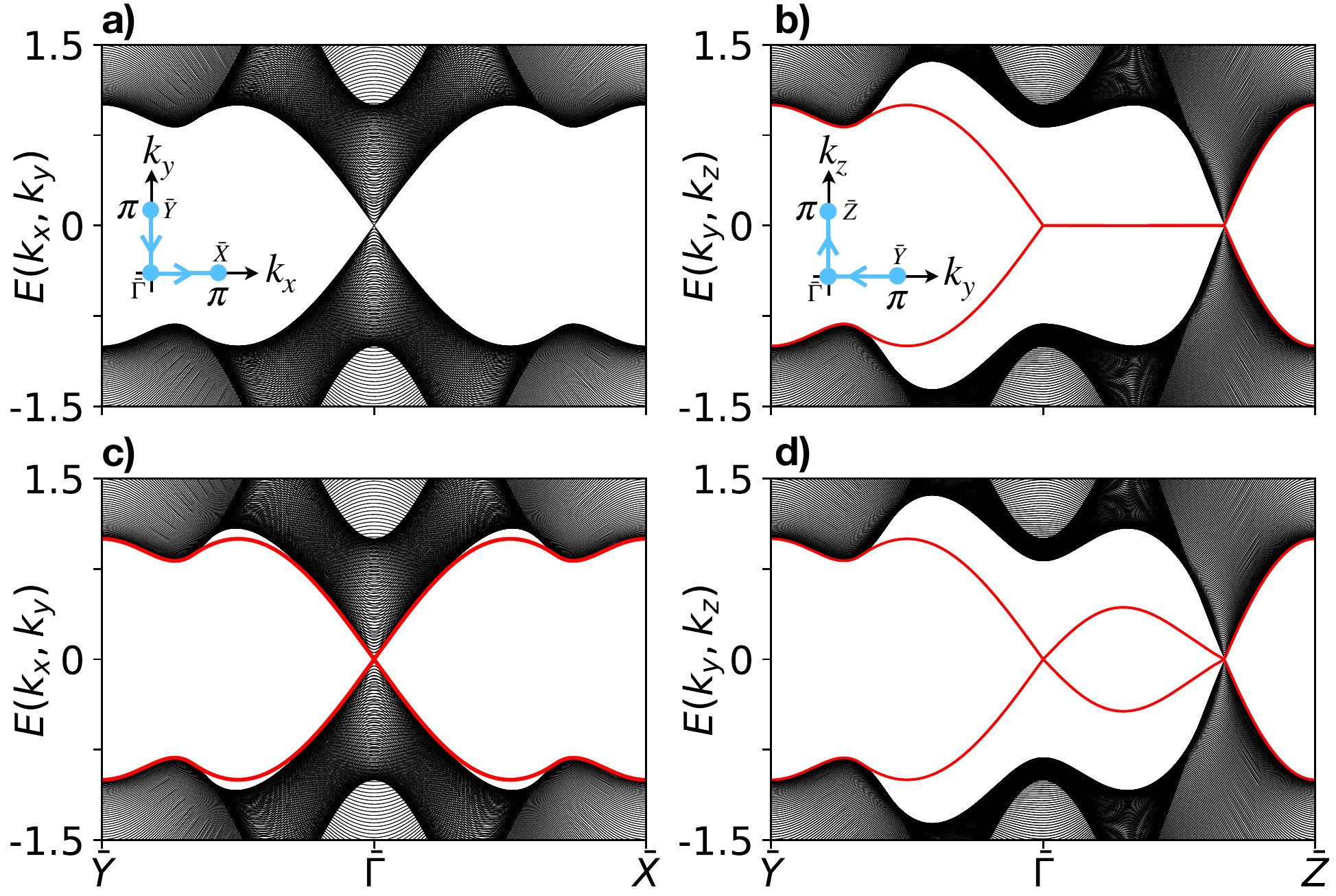}
\caption{(Color online) Surface band structure for the normal state. Common parameters are $m =3$, $t = t_z =2$, $\lambda=1$. $\eta=0$ in (a) and (b), and $\eta=0.5$ in (c) and (d). Surface Dirac cones are absent in (a) and (b)  and present in (c) and (d) at the center of the surface Brillouin zone. In each panel, the side with open boundary conditions is 200 lattice sites long.}
\label{normalstate}
\end{figure}

Usually, as the two symmetry-allowed $\eta$ terms in Eq.~(\ref{Hamiltonian}) only contribute cubic-order terms in momentum
to the continuum Dirac Hamiltonian,  they are neglected. While it is true that their higher-order contributions to the bulk 
can be safely neglected when focusing on the low-energy physics near the Dirac points, it has been demonstrated 
that their impact on the surface states, however, is significant \cite{Kargarian2016,Yan2020vortex,Le2018coexist}. Without the two $\eta$ terms ($\eta=0$), the DSM is found to 
harbor only Fermi arcs on the side surfaces. Remarkably, once  the two $\eta$ terms are present ($\eta\neq0$), the DSM 
 harbors not only Fermi arcs on the side surfaces, but also a single Dirac cone on each of the surfaces of
a cubic-geometry sample, resembling the surface Dirac cones in strong topological insulators. 
To have an intuitive 
picture of the qualitative difference between $\eta=0$ and $\eta\neq0$, we take 
$\{m,t,t_{z},\lambda\}=\{3,2,2,1\}$ so that the Dirac points are localized at $\bm{k}_{D;\pm}=\pm(0,0,2\pi/3)$
and then diagonalize 
the Hamiltonian in a cubic geometry with open boundary conditions in one direction and
periodic boundary conditions in the other two orthogonal directions. 
The corresponding energy spectra shown in Fig.~\ref{normalstate} clearly manifest the qualitative difference 
in surface states between $\eta=0$ and $\eta\neq0$. As we will show below, this remarkable difference 
will lead to distinct topological superconducting states. 

\section{Topological properties of superconducting Dirac semimetals}

Let us now focus on the superconducting state. 
Within the mean-field framework,  the Hamiltonian becomes $H=\frac{1}{2}\sum_{\bm{k}}\Psi_{\bm{k}}^{\dag}H_{\rm BdG}(\bm{k})\Psi_{\bm{k}}$, with 
$\Psi_{\bm{k}}^{\dag}=(\psi_{\bm{k}}^{\dag},\psi_{-\bm{k}})$ and the corresponding Bogoliubov-de Gennes (BdG)
Hamiltonian takes the form 
\begin{eqnarray}
H_{\rm BdG}(\bm{k})&=&\left(
                   \begin{array}{cc}
                     H_{\rm DSM}(\bm{k})-\mu & -is_{y}\Delta(\bm{k}) \\
                     is_{y}\Delta(\bm{k}) & \mu-H_{\rm DSM}^{*}(-\bm{k}) \\
                   \end{array}
                 \right),
\end{eqnarray}
where $\Delta(\bm{k})=\Delta_{0}-\Delta_{s}(\cos k_{x}+\cos k_{y})$ characterizes the $s_{\pm}$-wave pairing. Since the BdG Hamiltonian 
simultaneously has time-reversal symmetry and particle-hole symmetry, it belongs to the DIII class
in the ten-fold way classification~\cite{schnyder2008,kitaev2009}. Accordingly, its first-order topology is characterized 
by a winding number $N_{w}$ and follows a $\mathbb{Z}$ classification in three dimensions. When 
$N_{w}$ is a nonzero integer in a gapped superconductor, the bulk-boundary correspondence tells us that there are $N_{w}$ robust Majorana cones on 
an arbitrary surface~\cite{roy2008,qi2009b,volovik2009b}, irrespective of its orientation. 
A simple formula for $N_{w}$ valid in the weak-pairing limit is~\cite{qi2010d} 
\begin{eqnarray}
N_{w}=\frac{1}{2}\sum_{n}\text{sgn}(\Delta_{i})C_{1i}, 
\end{eqnarray}
where $C_{1i}$ denotes the first Chern number  and 
$\text{sgn}(\Delta_{i})$ denotes the sign of pairing on the $i$th Fermi surface. 
Since the simultaneous preservation of time-reversal symmetry and inversion symmetry 
forces $C_{1i}$ to vanish, $N_{w}$ thus identically vanishes, indicating 
that the first-order topology is always trivial for this Hamiltonian. Despite 
the absence of nontrivial first-order topology, the superconducting DSM, nevertheless, 
can be nontrivial in the higher-order topology and host interesting 
Majorana modes on the boundary.

\begin{figure}[t!]
\centering
\includegraphics[scale=0.232]{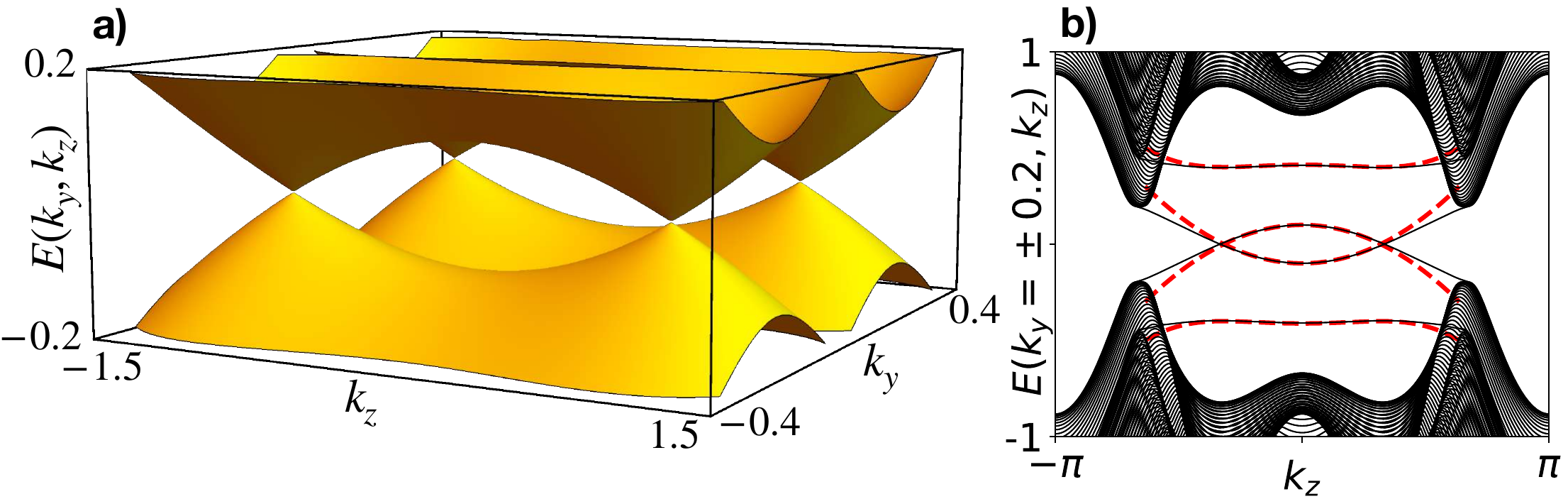}
\caption{(Color online) Chosen parameters are $m =3$, $t = t_z =2$, $\lambda=1$, $\eta=0$, $\mu = 0.2$, and $\Delta_0 = \Delta_s = 0.2$. Accordingly, $R_{\rm FS}=0.2$, $R_{\rm PNS}=\sqrt{2}$, and $R_{\rm BIS}=\sqrt{3}$. (a) The two middle bands of the surface Hamiltonian in Eq.~(\ref{surface}) touch
at $(k_{y},k_{z})=(\pm0.2,\pm1)$, forming four gapless Bogoliubov-Dirac  cones. (b) Energy spectrum along $k_{y}=\pm0.2$. The surface bands obtained from the low-energy analytical approach (midgap red dashed lines) agree excellently with those obtained by directly diagonalizing the full lattice Hamiltonian (midgap black solid lines, doubly degenerate) under open boundary conditions in the $x$ direction, confirming the existence of four gapless Bogoliubov-Dirac  cones on each of the two $x$-normal surfaces.  }\label{cone}
\end{figure}

The bulk spectrum of the superconducting DSM is gapped as long as the pairing node surface (PNS), which is the zero-value contour of 
$\Delta(\bm{k})$ in momentum space, does not cross the Fermi surface. 
On the boundary, the presence of superconductivity is also expected to 
gap out the topological surface states. An interesting question is whether it is possible that 
while the bulk states are fully gapped, the topological surface states are not fully gapped, so that 
there emerge certain types of gapless Bogoliubov quasiparticles on the boundary. 
We find that the answer is affirmative. To show this, the most intuitive approach 
is to derive the low-energy Hamiltonian for the surface states. Without loss of generality, we focus on the left $x$-normal surface and assume that the parameters $m$, $t$, and $t_z$ are chosen such that the BIS 
encloses the time-reversal invariant momentum $\boldsymbol{\Gamma}=(0,0,0)$. Following 
a  standard approach, we expand the lattice Hamiltonian 
around $\boldsymbol{\Gamma}$ to obtain the continuum bulk Hamiltonian and then 
find that the corresponding low-energy surface Hamiltonian takes the form (see Appendix~A)
\begin{eqnarray}
H_{s}(k_{y},k_{z})&\approx&\lambda k_{y}s_{z}+v_{z}(k_{y},k_{z})k_{z}\tau_{z}s_{y}-\mu\tau_{z}\nonumber\\
&&+\frac{\Delta_{s}}{2}\left(R_{\rm BIS}^{2}-R_{\rm PNS}^{2}
-\frac{t_{z}}{t}k_{z}^{2}\right)\tau_{y}s_{y},
\label{surface}
\end{eqnarray}
where $v_{z}(k_{y},k_{z})=-\eta(\tilde{m}+tk_{y}^{2}+t_{z}k_{z}^{2}/2)/t$ with $\tilde{m}=m-2t-t_{z}$, 
$R_{\rm BIS}=\sqrt{-2\tilde{m}/t}$, and $R_{\rm PNS}=\sqrt{-2\tilde{\Delta}/\Delta_{s}}$ 
with $\tilde{\Delta}=\Delta_{0}-2\Delta_{s}$. Here, we have already assumed 
$\{t,t_{z},\lambda,\Delta_{s}\}>0$ and $\{\tilde{m}, \tilde{\Delta}\}<0$.  According 
to the continuum bulk Hamiltonian, $R_{\rm BIS}$ and $R_{\rm PNS}$
correspond to the radii of BIS and PNS in the $k_{z}=0$ plane, respectively. It is worth noting that the 
surface states only exist in the regime satisfying 
$tk_{y}^{2}+t_{z}k_{z}^{2}<-2\tilde{m}$, which is just the projection of BIS
in the $k_{x}$ direction.

Let us first consider the $\eta=0$ case, where $v_{z}=0$ in this limit. Accordingly, 
the normal state has only Fermi arcs which are two straight lines at $k_{y}=\pm R_{\rm FS}$, 
where $R_{\rm FS}=|\mu/\lambda|$ corresponds to the maximum radius of the Fermi surface in the $k_{x}$-$k_{y}$ plane. 
The geometric meaning of this expression is that the Fermi arcs tangentially connect  with the projection 
of the Fermi surface in the surface Brillouin zone~\cite{Haldane2014}. 
Taking into account superconductivity, 
 we find from Eq.~(\ref{surface}) that the surface energy bands harbor
four cones with linear dispersion at $(k_{y},k_{z})=(\pm R_{\rm FS},\pm\sqrt{t(R_{\rm BIS}^{2}-R_{\rm PNS}^{2})/t_{z}})$
if $R_{\rm BIS}>R_{\rm PNS}$. As the Bogoliubov quasiparticle operators associated with these surface cones do not satisfy the self-conjugate property ($\gamma_{\bm{k}}^{\dag}=\gamma_{-\bm{k}}$ with $\gamma_{\bm{k}}^{\dag}$ denoting the quasiparticle creation operator at momentum $\bm{k}$), we dub them  Bogoliubov-Dirac cones to distinguish them from charge-neutral Majorana cones. 
Recalling that the precondition
for this result is $tk_{y}^{2}+t_{z}k_{z}^{2}<-2\tilde{m}$,
we find that the criterion for the existence of surface Bogoliubov-Dirac cones  needs to be 
modified as $R_{\rm FS}<R_{\rm PNS}<R_{\rm BIS}$.
This criterion corresponds to a simple geometric picture, namely, the PNS simultaneously encloses the bulk Fermi surface and intersects the BIS. In Fig.~\ref{cone}, we provide numerical results to show the existence 
of four gapless Bogoliubov-Dirac  cones on each of the side surfaces
(note that the system has $C_{4z}$-rotation symmetry) when the above-mentioned criterion  is fulfilled. Before proceeding to $\eta\neq0$, it is worth pointing out that every 
gapless Bogoliubov-Dirac cone has a topological protection due to 
the existence of chiral symmetry (the product of particle-hole symmetry and time-reversal symmetry) which will assign 
a topological winding number to characterize the band 
touching points of the surface energy spectrum~\cite{ryu2010} (see Appendix~A). 
Accordingly, one gapless Bogoliubov-Dirac cone can be gapped only when it meets another gapless Bogoliubov-Dirac cone characterized by 
an opposite winding number.

\begin{figure}[t!]
\centering
\includegraphics[scale=0.34]{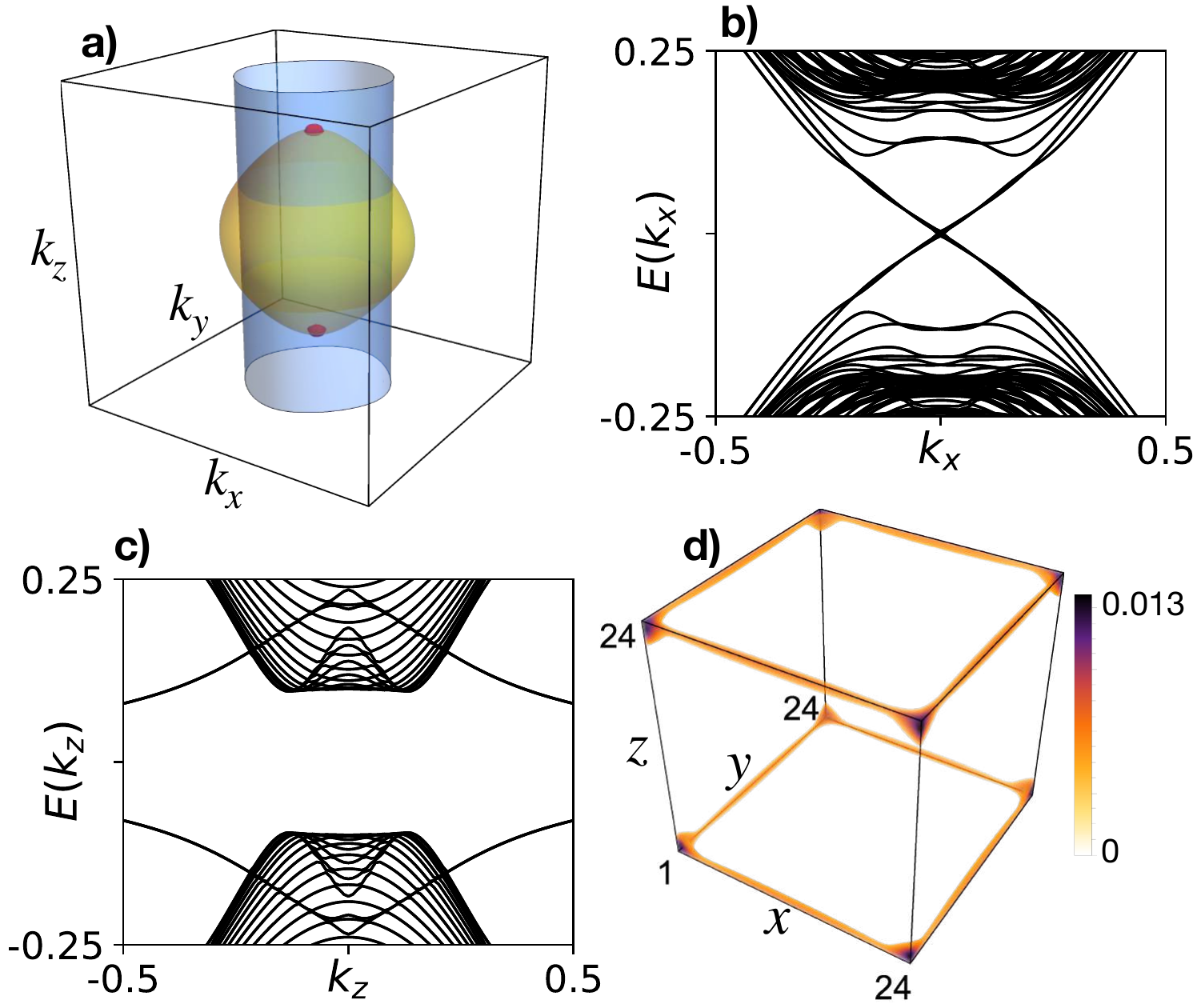}
\caption{(Color online) Chosen parameters are $m =3$, $t = t_z =2$, $\lambda=\eta=1$, $\mu = 0.2$, and $\Delta_0 = \Delta_s = 0.2$. (a) The PNS (blue), the BIS (yellow), and the two Fermi surfaces (red). (b) Helical Majorana hinge modes along the $x$ direction. Both $y$ and $z$ directions have open boundary conditions with $N_y = N_z = 60$ lattice sites. (c) No gapless hinge modes along the $z$ direction. Both $x$ and $y$ directions have open boundary conditions with $N_x = N_y = 50$. (d) The probability density distribution of the four eigenstates whose energies are in the middle of the BdG energy spectrum,  confirming the localization of the helical 
Majorana modes on the hinges.  All three directions have open boundary conditions with $N_x = N_y = N_z = 24$.  }\label{sotsc}
\end{figure}

Now we turn to the $\eta\neq0$ case for which surface Fermi arcs and Dirac cones coexist in the 
normal state. According to Eq.~(\ref{surface}), we find that if the energy spectrum harbors 
gapless Bogoliubov-Dirac  cones at $\eta=0$, these persist for nonzero $\eta$ as long as 
$|\eta|<\eta_{c}=\frac{2|\mu|}{R_{\rm PNS}^{2}}\sqrt{\frac{t_{z}}{t(R_{\rm BIS}^{2}-R_{\rm PNS}^{2})}}$. However, with 
the increase in $\eta$, the surface Bogoliubov-Dirac cones approach one another and become gapped pairwise when $|\eta|>\eta_{c}$, 
resulting in a fully gapped surface energy spectrum (see Appendix~A). Remarkably, after gapping out the
surface Bogoliubov-Dirac cones, we find that the superconductor becomes a  second-order 
time-reversal invariant TSC with helical Majorana hinge modes. To have an intuitive understanding 
of this transition, here we take the special case with $\mu=0$ for an analytical illustration. For this special case, 
$\eta_{c}=0$, suggesting that 
arbitrarily weak $\eta$ terms will gap out the surface Bogoliubov-Dirac cones. 
Focusing on the small momentum region, the surface Hamiltonian in 
Eq.~(\ref{surface}) can be simplified by 
neglecting the cubic-order momentum terms as 
\begin{align}
H_{s}(k_{y},k_{z})&=-\frac{\tilde{m}\eta}{t} k_{z}\tau_{z}s_{y}+\frac{\Delta_{s}}{2}(R_{\rm BIS}^{2}-R_{\rm PNS}^{2}
-\frac{t_{z}}{t}k_{z}^{2})\tau_{y}s_{y} \nonumber
\\&\quad
+\lambda k_{y}s_{z}.
\end{align}
When $\eta\neq0$ and $R_{\rm BIS}>R_{\rm PNS}$, the first line 
realizes a one-dimensional (1D) time-reversal invariant TSC in the $k_{z}$ direction~\cite{shen2013topological}. 
Considering a half-infinity surface occupying the region $0\leq z<+\infty$, 
doing the replacement $k_{z}\rightarrow-i\partial_z$
and solving the eigenvalue equation 
$H_{s}\phi_{\alpha}(z)=E_{\alpha}\phi_{\alpha}(z)$ under boundary conditions 
$\phi_{\alpha}(0)=\phi_{\alpha}(+\infty)=0$, one will find the existence of two 
branches of charge-neutral midgap states with opposite spin polarizations on the boundary of the surface, with their dispersions given 
by $E_{\alpha=1,2}=\pm \lambda k_{y}$ (see Appendix~A), indicating the appearance of helical Majorana modes on the hinges. 
In Fig.~\ref{sotsc}, we further provide numerical results for $\mu\neq0$ to support the realization of 
a second-order time-reversal invariant TSC with helical Majorana hinge modes
when the criterion established above is fulfilled.

\section{First-order time-reversal invariant topological superconductivity in thin-film superconducting Dirac semimetals}

For $s_{\pm}$-wave pairing, we have shown that the first-order topology is always trivial 
when time-reversal symmetry and inversion symmetry are preserved simultaneously. In the following, 
we consider reducing the bulk superconducting DSM to a thin film along the $z$ direction so that 
inversion symmetry can be easily broken by applying a gate voltage to the top and bottom layers~\cite{Zhang2021TRITSC}. 
Remarkably, we find that when $\eta\neq0$, a first-order time-reversal invariant TSC can be achieved (a discussion of the $\eta=0$ case is provided in Appendix~B). 
It is worth noting that although the thin-film superconducting  DSM still belongs to class DIII,
the classification of the gapped phases is changed from $\mathbb{Z}$ to $\mathbb{Z}_{2}$ due to the dimensional reduction, with
the $\mathbb{Z}_{2}$ invariant given in the weak-pairing limit by~\cite{qi2010d} 
\begin{eqnarray}
N_{\rm 2D}=\prod_{i}[\text{sgn}(\Delta_{i})]^{m_{i}}.\label{invariant}
\end{eqnarray}
Here, $m_{i}$ counts the number of time-reversal invariant momenta enclosed by the $i$th Fermi surface, 
and $N_{\rm 2D}=-1$ indicates the realization of a first-order time-reversal invariant TSC with helical Majorana edge modes~\cite{qi2009b,Nakosai2012,zhang2013kramers,wang2014TRI,Parhizgar2017,Casas2019,Zhang2021TRITSC}. 

\begin{figure}[t!]
\centering
\includegraphics[scale=0.292]{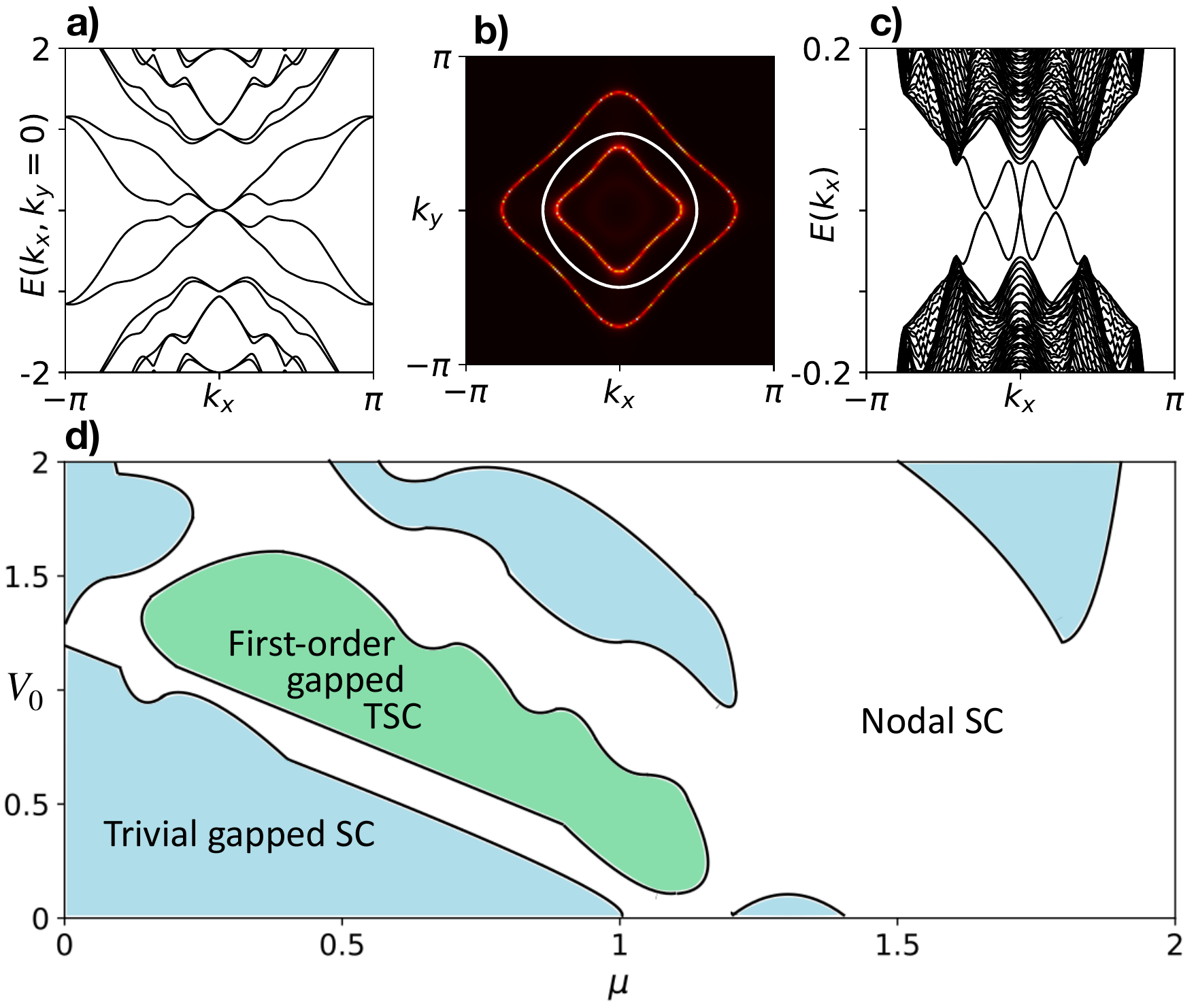}
\caption{(Color online) Chosen parameters are $m =3$, $t = t_z =2$, $\lambda=\eta=1$, and  $N_z = 5$. a) The normal-state energy spectrum along the $k_{y}=0$ line at $V_{0}=1$. b) The configurations of pairing line node (white) and Fermi surfaces (red) in the two-dimensional Brillouin zone for  $\Delta_0 = \Delta_s = 0.2$, $\mu = 0.7$
and $V_{0}=1$. c) With the same set of parameters as in b), the energy spectrum shows the existence of $\mathbb{Z}_{2}$ nontrivial helical Majorana modes in the gap when open boundary conditions are taken in both $y$ and $z$ directions with $N_y = 100$. The two tiny gaps at finite momentum are intrinsic and correspond to avoided crossings. d) The phase diagram includes three topologically distinct phases.} 
\label{fotsc}
\end{figure}

To be specific, here we consider the number of layers to be $N_{z}=5$ and add a potential profile 
of the form $V(z)\psi_{k_{x},k_{y},z}^{\dag}\psi_{k_{x},k_{y},z}$ to the BdG Hamiltonian, where $V(z)=V_{0}(2z-N_{z}-1)/(N_{z}-1)$  with 
$z=1,2,...,N_{z}$, {\it i.e.} the gate voltage varies linearly across the sample, so the voltage difference between top and bottom layers is $2V_{0}$. With the same set of parameters as 
in the bulk case, the corresponding normal-state energy spectra for the thin film 
are shown in Fig.~\ref{fotsc}(a). 
One finds that, in this case, the normal state is a two-dimensional semimetal with spin-split dispersion (away from time-reversal invariant momenta). Assuming the location 
of PNS to be fixed, we find that tuning the chemical potential can make the PNS fall between 
two disconnected Fermi surfaces, as shown in Fig.~\ref{fotsc}(b). In accordance with Eq.~(\ref{invariant}), 
it is readily found that $N_{\rm 2D}$ takes the nontrivial value $-1$ for the configuration in Fig.~\ref{fotsc}(b). 
By numerically calculating the energy spectra in a cylinder geometry, the existence of robust 
midgap helical Majorana edge modes confirms the realization of a first-order 
time-reversal invariant TSC, as shown in  Fig.~\ref{fotsc}(c). Moreover, the phase diagram
in Fig.~\ref{fotsc}(d) shows that, for a broad regime of $\mu$, the thin-film 
superconducting DSM can be made topologically nontrivial by tuning the gate voltage. 
\section{Discussion and conclusion}
We have uncovered topological criteria 
for the realization
of surface Bogoliubov-Dirac cones  and helical Majorana hinge modes in three-dimensional 
superconducting DSMs with $s_{\pm}$-wave pairing. 
Remarkably, the topological criteria admit a simple geometric
interpretation in terms of the relative configurations of BIS,
PNS, and Fermi surface. We have also shown that 
first-order time-reversal invariant TSCs can be realized in thin-film 
superconducting DSMs by applying a gate voltage to break inversion symmetry. 
Our work suggests that intrinsic superconductors
simultaneously hosting a gapless Dirac band structure and unconventional superconductivity
can realize a diversity of intrinsic time-reversal invariant TSCs and Majorana modes. Our predictions can be tested in iron-based superconductors such as  LiFe$_{1-x}$Co$_{x}$As~\cite{zhang2019multiple} 
by adjusting the doping level so as to position the Fermi energy near the bulk Dirac points. Experimentally, the surface Bogoliubov-Dirac cones can be detected by angle-resolved photoemission spectroscopy~\cite{zhang2018iron,zhang2019multiple}, and the helical Majorana modes can be measured by scanning tunneling microscopy~\cite{Kezilebieke2020} as well as contact methods \cite{Gray2019helical}.

\begin{acknowledgments}
M.Kh. and J.M. acknowledge support from NSERC Discovery Grant No.~RGPIN-2020-06999. J.M. also acknowledges support from NSERC Discovery Grant No.~RGPAS-2020-00064; the CRC Program; CIFAR; a Government of Alberta MIF Grant; a Tri-Agency NFRF Grant (Exploration Stream); and the PIMS CRG program. Z.-Y.Z. and Z.Y. are supported by the National Natural Science Foundation of China (Grants No.~11904417 and 
 12174455) and the Natural Science Foundation of Guangdong Province
 (Grant No.~2021B1515020026).
 \end{acknowledgments}
\begin{appendix}
\setcounter{figure}{0}
\renewcommand{\thefigure}{A\arabic{figure}}
\section{Derivation of the low-energy Hamiltonian for the surface states}
We start with the full  Bogoliubov-de Gennes (BdG) lattice Hamiltonian, which reads
\begin{align}
H_{\rm BdG}(\bm{k})&=[m-t(\cos k_{x}+\cos k_{y})-t_{z}\cos k_{z}]\tau_{z}\sigma_{z} \nonumber
\\&\quad 
+\lambda (\sin k_{x}s_{z}\sigma_{x}-\sin k_{y}\tau_{z}\sigma_{y}) -\mu\tau_{z} \nonumber
\\&\quad 
+\eta \sin k_{z}(\cos k_{x}-\cos k_{y})s_{x}\sigma_{x}
\nonumber
\\&\quad 
+2\eta \sin k_{x}\sin k_{y}\sin k_{z}\tau_{z}s_{y}\sigma_{x} \nonumber
\\&\quad 
+[\Delta_{0}-\Delta_{s}(\cos k_{x}+\cos k_{y})]\tau_{y}s_{y},
\end{align}
where the Pauli matrices $\sigma_{i}$, $s_{i}$, and $\tau_{i}$ act on the orbital,
spin, and particle-hole  degrees of freedom, respectively.
Similar to the main text, the lattice constants are set to unity   and the identity
matrices  are made implicit for brevity. To derive the low-energy Hamiltonian for the surface states, without loss of generality,
we consider that the band inversion surface (BIS) only encloses one time-reversal invariant momentum,
$\boldsymbol{\Gamma}=(0,0,0)$. Accordingly, we expand the lattice Hamiltonian around $\boldsymbol{\Gamma}$
to obtain the corresponding continuum bulk Hamiltonian,  which reads
\begin{align}
H_{c}(\bm{k})&=\left[\tilde{m}+\frac{t}{2}(k_{x}^{2}+k_{y}^{2})+\frac{t_{z}}{2}k_{z}^{2}\right]\tau_{z}\sigma_{z} \nonumber
\\&\quad
+\lambda( k_{x}s_{z}\sigma_{x}-  k_{y}\tau_{z}\sigma_{y}) -\mu\tau_{z} \nonumber
\\&\quad
-\frac{\eta}{2}k_{z}(k_{x}^{2}-k_{y}^{2})s_{x}\sigma_{x}
+2\eta k_{z}k_{x}k_{y}\tau_{z}s_{y}\sigma_{x}\nonumber
\\& \quad
+\left[\tilde{\Delta}+\frac{\Delta_{s}}{2}(k_{x}^{2}+k_{y}^{2})\right]\tau_{y}s_{y},
\end{align}
where $\tilde{m}=m-2t-t_{z}$ and $\tilde{\Delta}=\Delta_{0}-2\Delta_{s}$. Before proceeding, 
it is worth noting that while the low-energy bulk physics is dominated by the gapless bulk Dirac cones, 
one cannot use the low-energy bulk Hamiltonian expanded around the Dirac points to extract the low-energy 
boundary Hamiltonian describing  the surface states; instead, one needs to use 
the low-energy bulk Hamiltonian expanded around the band inversion momentum (above we 
have assumed it to be $\boldsymbol{\Gamma}$). 
The surface states originate from the band
inversion, so one should expand around the band inversion momentum to 
take into account the full band inversion region. On the other hand, 
the locations of Dirac points correspond to the boundary of the band inversion 
surface along the rotation-symmetric axis, so in fact one cannot obtain the
surface-state information through the low-energy Hamiltonian expanded 
around the Dirac points. In addition, it is also worth noting that 
we have only kept the leading term in momentum for each term in the continuum bulk Hamiltonian 
for simplicity. Such an approximation allows a simple analytic derivation of the low-energy 
boundary Hamiltonian, and it captures the essential physics quite accurately, particularly 
in the regime close to the band inversion momentum.

To be specific, in the following we assume $\{t,t_{z},\lambda,\eta\}$ to be all positive and
$\tilde{m}$ to be negative so that the normal state harbors a pair of Dirac points at $(0,0,\pm\sqrt{-2\tilde{m}/t_{z}})$.
For the pairing order parameter, we assume $\Delta_{s}>0$ but $\tilde{\Delta}<0$, so that the pairing amplitude has a nodal surface
in momentum space. For later discussion, we will introduce two quantities, $R_{\rm BIS}=\sqrt{-2\tilde{m}/t}$ and
$R_{\rm PNS}=\sqrt{-2\tilde{\Delta}/\Delta_{s}}$, which correspond to the radius of the ellipsoidal BIS
in the $k_{z}=0$ plane and the radius of the cylindrical pairing node surface (PNS), respectively. Geometrically, when
$0<R_{\rm PNS}<R_{\rm BIS}$, the BIS and PNS intersect.

We will focus on side surfaces which can harbor both Fermi arcs and Dirac cones.
Since the Hamiltonian has $C_{4z}$-rotation symmetry, we can just focus on
the $x$-normal surface. To be specific, we consider that the system occupies the region $0\leq x< +\infty$.
Since the presence of a boundary breaks the translation symmetry in the $x$ direction, $k_{x}$ needs  to be replaced
by $-i\partial_{x}$. Accordingly, we have
\begin{align}
\hspace{-2.5mm}
H_{c}(-i\partial_{x},k_{y},k_{z})&=\left(\tilde{m}-\frac{t}{2}\partial_{x}^{2}+\frac{t}{2}k_{y}^{2}+\frac{t_{z}}{2}k_{z}^{2}\right)\tau_{z}\sigma_{z} \nonumber
\\&\quad
-i\lambda\partial_{x}s_{z}\sigma_{x}-\lambda k_{y}\tau_{z}\sigma_{y}-\mu\tau_{z} \nonumber
\\&\quad
+\frac{\eta}{2}k_{z}(\partial_{x}^{2}+k_{y}^{2})s_{x}\sigma_{x}-2i\eta k_{z}k_{y}\partial_{x}\tau_{z}s_{y}\sigma_{x} \nonumber
\\&\quad
+\left(\tilde{\Delta}+\frac{\Delta_{s}}{2}k_{y}^{2}-\frac{\Delta_{s}}{2}\partial_{x}^{2}\right)\tau_{y}s_{y}.
\end{align}

In the next step, we decompose the Hamiltonian into two parts, {\it i.e.}, 
$H_{c}=H_{0}+H'$, with
\begin{widetext}
\begin{eqnarray}
H_{0}(-i\partial_{x},k_{y},k_{z})&=&(\tilde{m}-\frac{t}{2}\partial_{x}^{2}+\frac{t}{2}k_{y}^{2}+\frac{t_{z}}{2}k_{z}^{2})\tau_{z}\sigma_{z}
-i\lambda\partial_{x}s_{z}\sigma_{x}-\lambda k_{y}\tau_{z}\sigma_{y},\label{decompose}
\\
H^{'}(-i\partial_{x},k_{y},k_{z})&=&\frac{\eta}{2}k_{z}(\partial_{x}^{2}+k_{y}^{2})s_{x}\sigma_{x}
-2i\eta k_{z}k_{y}\partial_{x}\tau_{z}s_{y}\sigma_{x}-\mu\tau_{z}+(\tilde{\Delta}+\frac{\Delta_{s}}{2}k_{y}^{2}-\frac{\Delta_{s}}{2}\partial_{x}^{2})\tau_{y}s_{y},
\end{eqnarray}
\end{widetext}
where $H_{0}$ is the part describing the Dirac semimetal without the cubic-order terms. 
It is worth noting that we always put the terms with the same Pauli matrices together as they play
the same role. Moreover, as the pairing constants are much smaller than
the hopping constants in materials and as the regime in which the chemical potential
is close to the Dirac points is of particular interest, {\it i.e.}, $\mu\rightarrow0$,
we will treat all terms in $H^{'}$  as perturbations. In the following, we first solve the equation $H_{0}\psi_{\alpha}(x)=E_{\alpha}\psi_{\alpha}(x)$. For surface states
localized on  the $x=0$ surface, we demand that their wave functions satisfy the boundary conditions $\psi_{\alpha}(0)=\psi_{\alpha}(\infty)=0$. It is readily found that there are four solutions, with two solutions corresponding to $E_{\alpha}=\lambda k_{y}$ 
and the other two corresponding to $E_{\alpha}=-\lambda k_{y}$. The expressions for the four solutions can be 
compactly written as \cite{Yan2018hosc}
\begin{eqnarray}
\psi_{\alpha}=\mathcal{N}\sin(\kappa_{1}x) e^{-(\kappa_{2}x)}e^{ik_{y}y}e^{ik_{z}z}\chi_{\alpha},\label{eigenvector}
\end{eqnarray}
where the normalization constant is given by $\mathcal{N}=2\sqrt{\kappa_{2}(\kappa_{1}^{2}+\kappa_{2}^{2})/\kappa_{1}^{2}}$,  with
\begin{align}
\kappa_{1}&=\sqrt{\frac{-2\tilde{m}-tk_{y}^{2}-t_{z}k_{z}^{2}}{t}-\left(\frac{\lambda}{t}\right)^{2}},
\\
\kappa_{2}&=\frac{\lambda}{t}.
\end{align}
The spinor $\chi_{\alpha}$ satisfies $\tau_{z}s_{z}\sigma_{y}\chi_{\alpha}=-\chi_{\alpha}$. Here, without loss of generality, we choose
$\chi_{1}=|\tau_{z}=1,s_{z}=1,\sigma_{y}=-1\rangle$, $\chi_{2}=|\tau_{z}=1,s_{z}=-1,\sigma_{y}=1\rangle$,
$\chi_{3}=|\tau_{z}=-1,s_{z}=1,\sigma_{y}=1\rangle$ and $\chi_{4}=|\tau_{z}=-1,s_{z}=-1,\sigma_{y}=-1\rangle$.
The normalization of the wave functions suggests that the boundary modes exist only when $(\kappa_{1}^{2}+\kappa_{2}^{2})>0$,
{\it i.e.} $tk_{y}^{2}+t_{z}k_{z}^{2}<-2\tilde{m}$, which is just the projection of BIS
in the $k_{x}$ direction. In the basis $(\psi_{1},\psi_{2},\psi_{3},\psi_{4})^{T}$, the surface-state 
Hamiltonian contributed by $H_{0}$ reads
\begin{eqnarray}
H_{s}^{(0)}(k_{y},k_{z})=\lambda k_{y}s_{z},
\end{eqnarray}
which only leads to straight Fermi arcs. For notational simplicity, we still make 
the identity matrix implicit. Taking into account $H'$, its contribution can be 
determined by the standard perturbation theory, 
\begin{eqnarray}
\hspace{-9mm}
[H_{s}'(k_{y},k_{z})]_{\alpha\beta}=\int_{0}^{\infty}
\hspace{-3mm}
\psi_{\alpha}^{\dag}(x)H'(-i\partial_{x},k_{y},k_{z})\psi_{\beta}(x)dx.
\end{eqnarray}
In terms of the Pauli matrices, one finds 
\begin{align}
H_{s}'(k_{y},k_{z}) &= v_{z}(k_{y},k_{z})k_{z}\tau_{z}s_{y}-\mu\tau_{z} \nonumber
\\& \quad
+(\tilde{\Delta}-\frac{\Delta_{s}\tilde{m}}{t}
-\frac{\Delta_{s}t_{z}}{2t}k_{z}^{2})\tau_{y}s_{y}
\end{align}
with \cite{Yan2020vortex}
\begin{widetext}
\begin{equation}
\begin{split}
v_{z}(k_{y},k_{z}) &= -\mathcal{N}^{2}\int_{0}^{\infty}\sin(\kappa_{1}x) e^{-(\kappa_{2}x)}\frac{\eta}{2}(\partial_{x}^{2}+k_{y}^{2})\sin(\kappa_{1}x) e^{-(\kappa_{2}x)}dx
\\
&= -\frac{\eta}{t}(\tilde{m}+tk_{y}^{2}+t_{z}k_{z}^{2}/2).
\end{split}
\end{equation}
\end{widetext}
Putting the two parts together, the low-energy Hamiltonian describing 
the surface states on the $x=0$ surface has the form
\begin{align}
H_{s}(k_{y},k_{z})& = H_{s}^{0}+H_{s}'\nonumber
\\&
=\lambda k_{y}s_{z}+v_{z}(k_{y},k_{z})k_{z}\tau_{z}s_{y}-\mu\tau_{z}
\nonumber
\\&\quad
+(\tilde{\Delta}-\frac{\Delta_{s}\tilde{m}}{t}
-\frac{\Delta_{s}t_{z}}{2t}k_{z}^{2})\tau_{y}s_{y}.
\end{align}
It is readily found that the boundary Hamiltonian
preserves all nonspatial symmetries of the bulk Hamiltonian, including
the time-reversal symmetry ($\mathcal{T}=is_{y}\mathcal{K}$),
particle-hole symmetry ($\mathcal{P}=\tau_{x}\mathcal{K}$), and their combination, the chiral
symmetry ($\mathcal{C}=\tau_{x}s_{y}$).

Next, let us rewrite the Hamiltonian as
\begin{align}
\hspace{-2mm}
H_{s}(k_{y},k_{z})&=\lambda k_{y}s_{z}+v_{z}(k_{y},k_{z})k_{z}\tau_{z}s_{y}-\mu\tau_{z}
\nonumber
\\&\quad
+\frac{\Delta_{s}}{2}\left(\frac{2\tilde{\Delta}}{\Delta_{s}}-\frac{2\tilde{m}}{t}
-\frac{t_{z}}{t}k_{z}^{2}\right)\tau_{y}s_{y}\nonumber
\\&
=\lambda k_{y}s_{z} + v_{z}(k_{y},k_{z})k_{z}\tau_{z}s_{y} -\mu\tau_{z} \nonumber
\\&\quad
+\frac{\Delta_{s}}{2}\left(R_{\rm BIS}^{2}-R_{\rm PNS}^{2}
-\frac{t_{z}}{t}k_{z}^{2}\right)\tau_{y}s_{y}, \label{effective}
\end{align}
which is Eq.~(\ref{surface}). When $\eta=0$, the Hamiltonian reduces to
\begin{align}
\hspace{-3mm}
H_{s}(k_{y},k_{z})&=\lambda k_{y}s_{z}-\mu\tau_{z} \nonumber
\\& \quad
+\frac{\Delta_{s}}{2}\left(R_{\rm BIS}^{2}-R_{\rm PNS}^{2}
-\frac{t_{z}}{t}k_{z}^{2}\right)\tau_{y}s_{y}.
\end{align}
At $\mu=0$, one can find that there are two cones with linear dispersion
and double degeneracy at $(k_{y},k_{z})=\left(0,\pm\sqrt{\frac{t}{t_{z}}(R_{\rm BIS}^{2}-R_{\rm PNS}^{2})}\right)$. Since the Hamiltonian can be decomposed into 
two decoupled parts when $\mu=0$, it is easy to see that the Bogoliubov quasiparticle operators 
will take the form $\gamma_{\bm{k},1}=u_{\bm{k},1}c_{\bm{k},\uparrow}+v_{\bm{k},1}c_{-\bm{k},\downarrow}^{\dag}$ or $\gamma_{\bm{k},2}=u_{\bm{k},2}c_{\bm{k},\downarrow}+v_{\bm{k},2}c_{-\bm{k},\uparrow}^{\dag}$ (the concrete expressions for $u_{\bm{k},1(2)}$ and $v_{\bm{k},1(2)}$ are not important here). In each case, the quasiparticle operators do not satisfy the 
self-conjugate property $\gamma_{\bm{k},1(2)}\neq\gamma_{-\bm{k},1(2)}^{\dag}$ as the electron part and hole part have opposite spin polarizations.  
Therefore we dub these cones with linear dispersion as Bogoliubov-Dirac cones to distinguish them from  Majorana cones.
 Recall that the gapless surface states only exist within the regime satisfying
$tk_{y}^{2}+t_{z}k_{z}^{2}<-2\tilde{m}$, {\it i.e.} $k_{y}^{2}+\frac{t_{z}}{t}k_{z}^{2}<R^{2}_{\rm BIS}$.
Therefore, the condition for the existence of Bogoliubov-Dirac cones at $\mu=0$ is very simple.
That is, $0<R_{\rm PNS}<R_{\rm BIS}$. Geometrically, this corresponds to the BIS and PNS intersecting in momentum space. Once $\mu\neq0$, the double degeneracy
of the Bogoliubov-Dirac cones at $\mu=0$ is split, and there are four separated  Bogoliubov-Dirac cones, with their locations being at
$(k_{y},k_{z})=(\pm\mu/\lambda,\pm\sqrt{\frac{t}{t_{z}}(R_{\rm BIS}^{2}-R_{\rm PNS}^{2})})$.
Since the Bogoliubov-Dirac cones must exist in the regime satisfying $k_{y}^{2}+\frac{t_{z}}{t}k_{z}^{2}<R^{2}_{\rm BIS}$,
the condition for their existence becomes $|\frac{\mu}{\lambda}|<R_{\rm PNS}<R_{\rm BIS}$.
Interestingly, $|\frac{\mu}{\lambda}|$ also has a geometric interpretation. To see this, let us focus on
the normal state and investigate the bulk Fermi surface. When $\eta=0$, the energy spectrum 
for the normal state is 
\begin{eqnarray}
E(\bm{k})=\pm\sqrt{\lambda^{2}(k_{x}^{2}+k_{y}^{2})+M^{2}(\bm{k})}
\end{eqnarray}
where $M(\bm{k})=\tilde{m}+\frac{t}{2}(k_{x}^{2}+k_{y}^{2})+\frac{t_{z}}{2}k_{z}^{2}$. The bulk Fermi surface is determined by 
\begin{eqnarray}
\lambda^{2}(k_{x}^{2}+k_{y}^{2})+M^{2}(\bm{k})=\mu^{2}
\end{eqnarray}
It is readily found that the maximum radius of the Fermi surface in the $k_{x}$-$k_{y}$ plane is
equal to $|\frac{\mu}{\lambda}|$. Defining $R_{\rm FS}=|\frac{\mu}{\lambda}|$, the criterion
for the existence of surface Bogoliubov-Dirac cones can be rewritten as $R_{\rm FS}<R_{\rm PNS}<R_{\rm BIS}$.
This form describes a very simple geometric picture. That is, the PNS encloses the bulk Fermi
surface and simultaneously intersects the BIS.
Before ending this part, let us further give a discussion 
of the topological protection of the surface Bogoliubov-Dirac cones. As we mentioned above, the 
two-dimensional boundary 
Hamiltonian inherits the chiral symmetry from the three-dimensional bulk. Due to the existence of 
chiral symmetry, the band touching points of the surface energy spectrum can be assigned 
a  winding number to characterize their topology.
First, one can change the basis so that the chiral operator takes a diagonal form in the new basis. 
Accordingly, it is known that the Hamiltonian will become off-diagonal,  
with the form 
\begin{eqnarray}
\tilde{H}(k_{y},k_{z})=\left(
                               \begin{array}{cc}
                                 0 & Q(k_{y},k_{z}) \\
                                 Q^{\dag}(k_{y},k_{z}) & 0 \\
                               \end{array}
                             \right), 
\end{eqnarray}
where $Q(k_{y},k_{z})$ is a $2\times2$ matrix, with its elements $Q_{11}=Q_{22}=\lambda k_{y}$, $Q_{12}=i\mu-iv_{z}(k_{y},k_{z})k_{z}+\frac{\Delta_{s}}{2}(R_{\rm BIS}^{2}-R_{\rm PNS}^{2}
-\frac{t_{z}}{t}k_{z}^{2})$, and $Q_{21}=-i\mu-iv_{z}(k_{y},k_{z})k_{z}-\frac{\Delta_{s}}{2}(R_{\rm BIS}^{2}-R_{\rm PNS}^{2}
-\frac{t_{z}}{t}k_{z}^{2})$. When a closed path is chosen to enclose one band touching 
point of the surface energy spectrum, a winding number can be defined to characterize the band touching point
in accordance with the below formula:~\cite{ryu2010}
\begin{eqnarray}
\omega=\frac{i}{2\pi}\oint_{C}\text{Tr}[Q^{-1}\partial_{k}Q]dk.
\end{eqnarray}
The topological nature of the winding number guarantees the robustness of separated band 
touching points. As a result, one gapless Bogoliubov-Dirac cone can be gapped only when it meets another gapless Bogoliubov-Dirac cone characterized by an opposite winding number. 

\begin{figure*}[t!]
\centering
\includegraphics[scale=0.55]{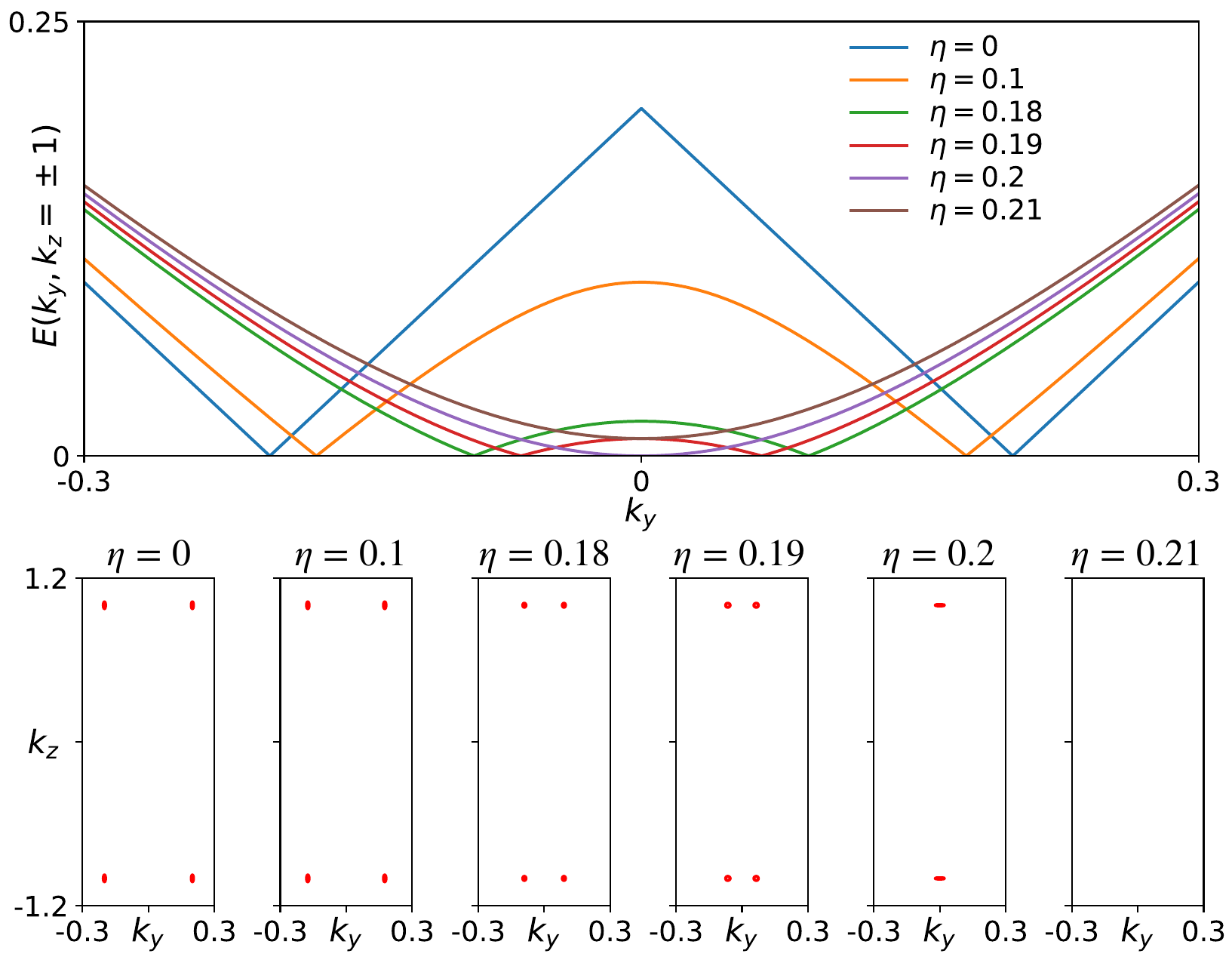}
\caption{(Color online) Chosen parameters are $m =3$, $t = t_z =2$, $\lambda=1$, $\mu=0.2$, and $\Delta_0 =\Delta_s = 0.2$.
Accordingly, $R_{\rm BIS}=\sqrt{3}$,  $R_{\rm PNS}=\sqrt{2}$, and $\eta_{c}=0.2$. The $x$-normal surface band structure at $k_z = \pm 1$ given by Eq.~(\ref{effective}) (top panel) and the position of four surface Bogoliubov-Dirac cones (bottom panels) for different values of $\eta$. As expected, only the $k_y$ coordinate of the position of the surface Bogoliubov-Dirac cones depends on $\eta$. Pairwise annihilation of the cones occurs at the critical value $\eta=0.2$. }\label{analytic}
\end{figure*}

\begin{figure*}[t!]
\centering
\includegraphics[scale=0.8]{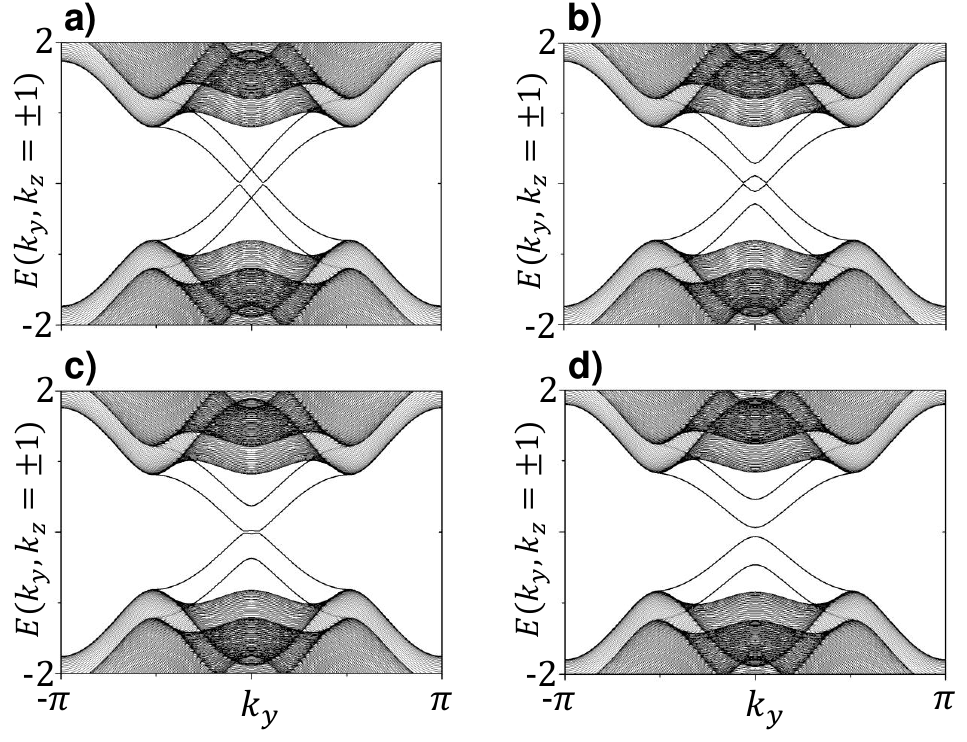}
\caption{Energy spectrum of the full lattice Hamiltonian under open boundary conditions in the $x$ direction and periodic boundary conditions in the $y$ and $z$ directions. Chosen parameters are $m =3$, $t = t_z =2$, $\lambda=1$, $\mu=0.2$, and $\Delta_0 =\Delta_s = 0.2$. a) $\eta=0$, b) $\eta=0.1$, c) $\eta=0.2$, d) $\eta=0.3$. }\label{numerical}
\end{figure*}

Now let us consider $\eta\neq0$. Accordingly, the surface energy spectrum becomes
\begin{widetext}
\begin{eqnarray}
E(k_{y},k_{z})=\pm\sqrt{\left(\sqrt{\lambda^{2}k_{y}^{2}+v_{z}^{2}(k_{y},k_{z})k_{z}^{2}}\pm\mu\right)^{2}+\frac{\Delta_{s}^2}{4}\left(R_{\rm BIS}^{2}-R_{\rm PNS}^{2}
-\frac{t_{z}}{t}k_{z}^{2}\right)^{2}}.
\end{eqnarray}
\end{widetext}
The surface Bogoliubov-Dirac cones, if they remain, are located at a value of $k_{z}=\pm\sqrt{t(R_{\rm BIS}^{2}-R_{\rm PNS}^{2})/t_{z}}$ independent of $\eta$. The $k_{y}$ value needs to be determined by
solving the equation
\begin{eqnarray}
\hspace{-8mm}
\lambda^{2}k_{y}^{2}+\eta^{2}(k_{y}^{2}+\tilde{\Delta}/\Delta_{s})^{2}t(R_{\rm BIS}^{2}-R_{\rm PNS}^{2})/t_{z}=\mu^{2},
\end{eqnarray}
or in the standard form
\begin{align}
&[\eta^{2}t(R_{\rm BIS}^{2}-R_{\rm PNS}^{2})/t_{z}] k_{y}^{4} \nonumber
\\&
+[\lambda^{2}+2\eta^{2}(\tilde{\Delta}/\Delta_{s})t(R_{\rm BIS}^{2}-R_{\rm PNS}^{2})/t_{z}]k_{y}^{2} \nonumber
\\&
+\eta^{2}(\tilde{\Delta}/\Delta_{s})^{2}t(R_{\rm BIS}^{2}-R_{\rm PNS}^{2})/t_{z}-\mu^{2}=0.
\end{align}
By defining 
\begin{align}
a &\equiv[\eta^{2}t(R_{\rm BIS}^{2}-R_{\rm PNS}^{2})/t_{z}],
\\
b &\equiv[\lambda^{2}+2\eta^{2}(\tilde{\Delta}/\Delta_{s})t(R_{\rm BIS}^{2}-R_{\rm PNS}^{2})/t_{z}],
\\
c& \equiv\eta^{2}(\tilde{\Delta}/\Delta_{s})^{2}t(R_{\rm BIS}^{2}-R_{\rm PNS}^{2})/t_{z}-\mu^{2},
\end{align}
it is known that the solutions for $k_{y}^{2}$ take the standard form 
\begin{eqnarray}
k_{y}^{2}=\frac{-b\pm\sqrt{b^{2}-4ac}}{2a}.
\end{eqnarray}
There will exist gapless Bogoliubov-Dirac cones in the surface Brillouin zone as long as 
real and positive solutions for $k_{y}^{2}$ exist. As we are interested in the movements
of the Bogoliubov-Dirac cones with the increase in $\eta$ from $0$, in the following we focus 
on the case with $b>0$ to give a discussion. As here the parameter $a$ is positive, 
the existence of a physical solution then requires $c<0$. Accordingly, 
one can find that the condition for the existence of gapless Bogoliubov-Dirac cones 
is 
\begin{align}
|\eta|<\eta_{c}&=\frac{|\mu|\Delta_{s}}{|\tilde{\Delta}|}\sqrt{\frac{t_{z}}{t(R_{\rm BIS}^{2}-R_{\rm PNS}^{2})}}
\nonumber
\\
&=\frac{2|\mu|}{R_{\rm PNS}^{2}}\sqrt{\frac{t_{z}}{t(R_{\rm BIS}^{2}-R_{\rm PNS}^{2})}}.
\label{formula}
\end{align}
Putting $\eta_{c}$ back into the formula for $b$, one obtains 
\begin{eqnarray}
b=\lambda^{2}+2\mu^{2}\frac{\Delta_{s}}{\tilde{\Delta}}.
\end{eqnarray}
As long as the chemical potential $|\mu|<\mu_{c}=\sqrt{|\frac{\tilde{\Delta}}{2\Delta_{s}}|}\lambda$, the parameter $b$
is positive, and the above formula for $\eta_{c}$ is valid. To intuitively see the effect of $\eta$ terms on $k_{y}^{2}$, we consider $|\mu|<\mu_{c}$ and $\eta$ to be small so that we can do
an expansion in $\eta$. To second order, we find
\begin{widetext}
\begin{eqnarray}
k_{y}^{2}\approx\frac{\mu^{2}}{\lambda^{2}}-\frac{\eta^{2}}{\lambda^{2}}\frac{t}{t_{z}}(R_{\rm BIS}^{2}-R_{\rm PNS}^{2})
\left(\frac{\tilde{\Delta}}{\Delta_{s}}\right)^{2}\left[1+\frac{\mu^{2}}{\lambda^{2}}\frac{\Delta_{s}}{\tilde{\Delta}}\right].
\end{eqnarray}
\end{widetext}

In the weakly doped regime, $\mu\ll\lambda$, one can see that the $\eta$ terms
decrease the separation of surface Bogoliubov-Dirac cones in the $k_{y}$ direction,
consistent with the picture that the surface Bogoliubov-Dirac cones will annihilate each other
when $\eta$ is larger than a critical value.
In Fig.~\ref{analytic}, we show the evolution of the positions
of surface Bogoliubov-Dirac cones with respect to $\eta$ explicitly. According to
this evolution, one can find that the value at which the surface Bogoliubov-Dirac
cones merge in pairs  agrees with the formula for $\eta_{c}$ in Eq.~(\ref{formula}).
By diagonalizing the full lattice Hamiltonian with open boundary conditions in
the $x$ direction, we find that the locations and evolution of surface Bogoliubov-Dirac cones
on the $x$-normal surface agree well with the analytical analysis above, as shown in Fig.~\ref{numerical}.

After gapping out the surface Bogoliubov-Dirac cones, we have shown both analytically and numerically that one-dimensional propagating helical Majorana modes will emerge on the hinges 
of a cubic sample. Here, we provide more details about the analytical derivation of the low-energy 
Hamiltonian for the helical Majorana hinge modes at the limit $\mu=0$. At $\mu=0$, 
the surface-state Hamiltonian becomes 
\begin{align}
H_{s}(k_{y},k_{z}) &=\lambda k_{y}s_{z}-\frac{\eta k_{z}}{t}(\tilde{m}+tk_{y}^{2}+t_{z}k_{z}^{2}/2)\tau_{z}s_{y} \nonumber
\\ &\quad
+\frac{\Delta_{s}}{2}(R_{\rm BIS}^{2}-R_{\rm PNS}^{2}
-\frac{t_{z}}{t}k_{z}^{2})\tau_{y}s_{y}.
\end{align}
It is readily found that the energy spectrum for this Hamiltonian is fully gapped as long as $\eta\neq0$ and 
$R_{\rm BIS}\neq R_{\rm PNS}$. Let us focus on the small-momentum region; accordingly, we will only keep the leading momentum terms in each term of the surface-state Hamiltonian. Then the Hamiltonian reduces to
\begin{align}
H_{s}(k_{y},k_{z}) &= \lambda k_{y}s_{z}-\frac{\eta \tilde{m}k_{z}}{t}\tau_{z}s_{y} \nonumber
\\& \quad
+\frac{\Delta_{s}}{2}(R_{\rm BIS}^{2}-R_{\rm PNS}^{2}
-\frac{t_{z}}{t}k_{z}^{2})\tau_{y}s_{y}.
\end{align}
If the open boundary condition is further taken in the $z$ direction, 
then the Hamiltonian becomes 
\begin{align}
\hspace{-3mm}
H_{s}(k_{y},-i\partial_{z})
&=\lambda k_{y}s_{z}+i\frac{\eta \tilde{m}}{t}\tau_{z}s_{y}\partial_{z}
\nonumber
\\& \quad
+\frac{\Delta_{s}}{2}(R_{\rm BIS}^{2}-R_{\rm PNS}^{2}
+\frac{t_{z}}{t}\partial_{z}^{2})\tau_{y}s_{y}.
\end{align}
As the Hamiltonian takes a form similar to $H_{0}$ in Eq.~ (\ref{decompose}), 
one can easily find that if we consider a half-infinity sample with the boundary 
at $z=0$ (the boundary is in fact a hinge as it corresponds to the boundary of a surface), 
there exist two solutions satisfying the eigenvalue equation 
$H_{s}\phi_{\alpha}(z)=E_{\alpha}\phi_{\alpha}(z)$
and the boundary condition $\phi_{\alpha}(0)=\phi_{\alpha}(\infty)=0$. 
The expressions for the two solutions are similar to those in Eq.~(\ref{eigenvector}), 
\begin{eqnarray}
\phi_{\alpha}(z)=\mathcal{N}'\sin(\kappa'_{1}z) e^{-(\kappa'_{2}z)}e^{ik_{y}y}\chi_{\alpha}',
\end{eqnarray}
where the normalization constant is given by $\mathcal{N}'=2\sqrt{\kappa_{2}'(\kappa_{1}^{'2}+\kappa_{2}^{'2})/\kappa_{1}^{'2}}$,
with
\begin{align}
\kappa_{1}' &=\sqrt{\frac{t(R_{\rm BIS}^{2}-R_{\rm PNS}^{2})}{t_{z}}-\left(\frac{\eta\tilde{m}}{t_{z}\Delta_{s}}\right)^{2}},
\\
\kappa_{2}' &=-\frac{\eta\tilde{m}}{t_{z}\Delta_{s}}.
\end{align}
The normalization of the wave functions requires $R_{\rm BIS}^{2}>R_{\rm PNS}^{2}$, indicating 
that the crossing of the BIS and PNS is a precondition for the realization of the helical Majorana hinge modes. 
Here the two spinors $\chi_{\alpha}'$ can be chosen as  $\chi_{1}'=|\tau_{x}=1,s_{z}=1\rangle$ and
$\chi_{2}'=|\tau_{x}=1,s_{z}=-1\rangle$. Correspondingly, $E_{1}=\lambda k_{y}$ and $E_{2}=-\lambda k_{y}$. As the two spinors indicate that 
each branch of the hinge states is spin-polarized and 
an equal superposition of electron and hole, this analysis 
confirms that the two branches of hinge states correspond to 
a pair of helical Majorana modes. In the basis $(\phi_{1},\phi_{2})^{T}$, the 
low-energy Hamiltonian that describes the helical Majorana hinge modes 
reads
\begin{eqnarray}
H_{h}(k_{y})=\lambda k_{y}s_{z}.
\end{eqnarray}
\section{Importance of $\eta$ terms for the realization of first-order time-reversal invariant topological superconductivity in thin films of superconducting Dirac semimetal}
\setcounter{figure}{0}
\renewcommand{\thefigure}{B\arabic{figure}}
\begin{figure*}[t!]
\centering
\includegraphics[scale=0.55]{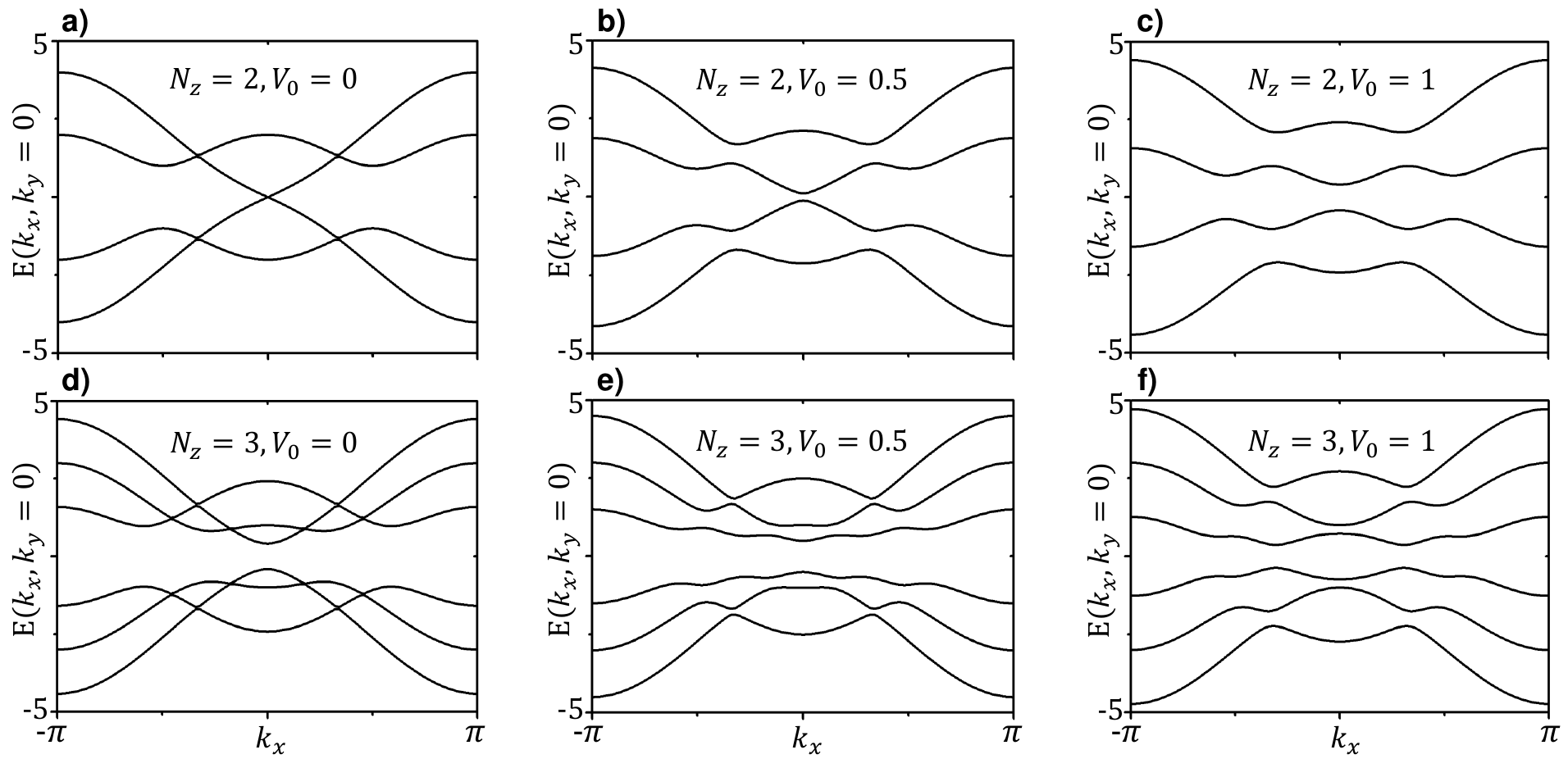}
\caption{ The evolution of normal-state energy spectra with respect to gate potential for bilayer and trilayer thin films.  Chosen parameters are $m =3$, $t = t_z =2$, $\lambda=1$, and $\eta=0$. The gate voltage cannot lift the spin degeneracy when the $\eta$ terms are absent. }\label{zeroeta}
\end{figure*}

\begin{figure*}[t!]
\centering
\includegraphics[scale=0.55]{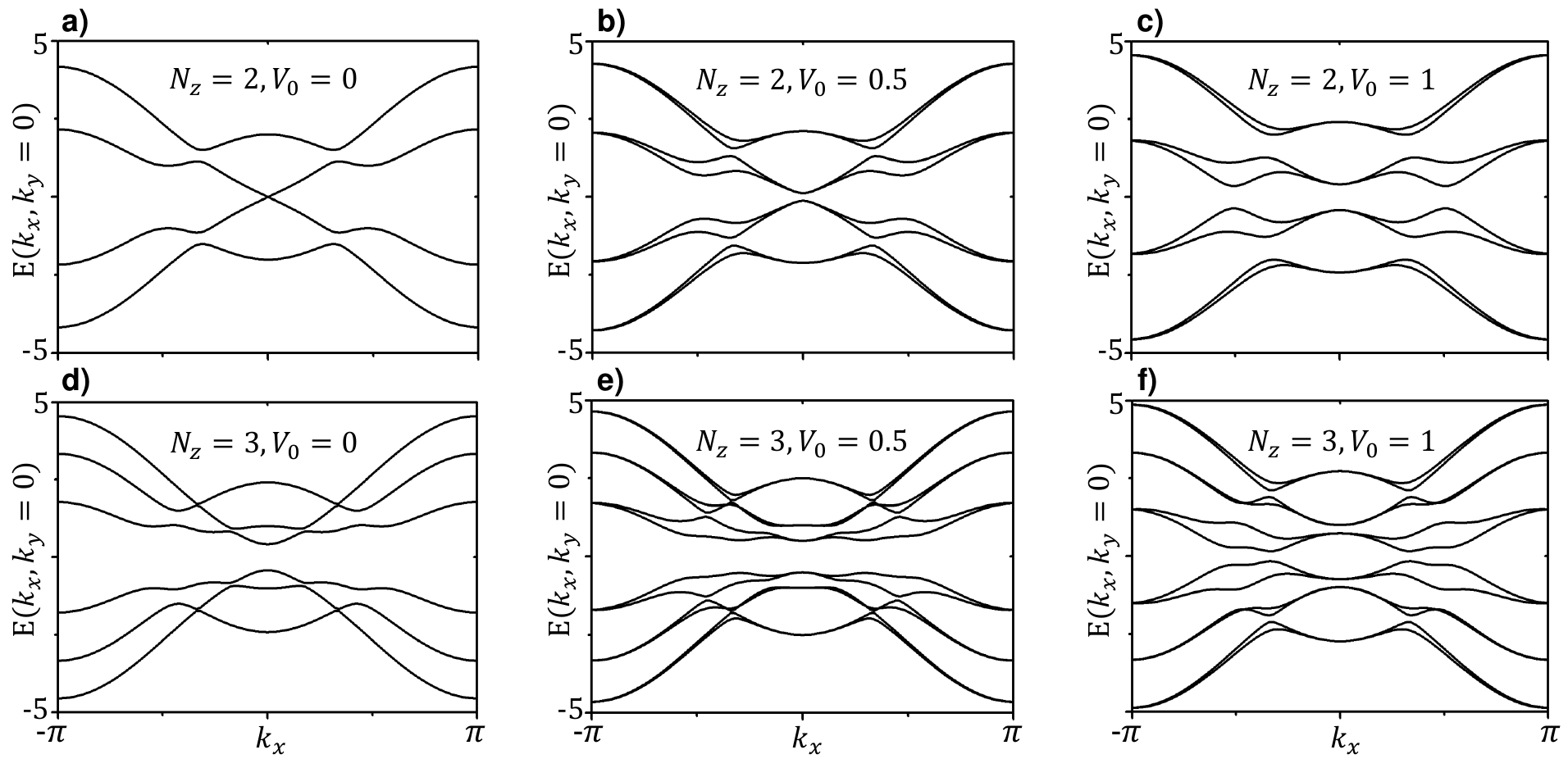}
\caption{ The evolution of normal-state energy spectra with respect to gate potential for bilayer and trilayer thin films. Chosen parameters are $m =3$, $t = t_z =2$, $\lambda=1$, and $\eta=1$. The gate voltage lifts the spin degeneracy when the $\eta$ terms are finite.}
\label{finiteeta}
\end{figure*}

In this appendix, we will show that the $\eta$ terms are also crucial for the realization of first-order
time-reversal invariant topological superconductivity in thin films of the superconducting Dirac semimetal. Before proceeding, we recall the fact that,
for the even-parity pairing discussed here, lifting the spin degeneracy of the Fermi surface is a precondition
for the realization of first-order time-reversal invariant topological superconductivity in two dimensions.

We first investigate the energy spectrum of thin-film Dirac semimetals when $\eta=0$ and superconductivity
is absent.  To be specific, here we focus on thin films with number of layers $N_{z}=2$ and $N_{z}=3$.
We find that, for both the bilayer and trilayer,
while the gate voltage can strongly modify the dispersions of the energy bands, it cannot lift
the spin degeneracy,  as shown in Fig.~\ref{zeroeta}. Since the double degeneracy of the energy bands
cannot be lifted by the gate voltage, this suggests that when the $\eta$ terms are absent,
the naive approach of using gate voltage to drive the superconducting Dirac semimetal with even-parity pairing
into a first-order
time-reversal invariant topological superconductor does not work.

For comparison, we change $\eta$ from $0$ to $1$ and keep other parameters fixed, with the corresponding
energy bands shown in Fig.~\ref{finiteeta}. One can see that, for both the bilayer and the trilayer thin films,
the double degeneracy of energy bands is lifted by a finite gate voltage, which makes the realization
of first-order time-reversal invariant topological superconductivity possible.

To understand the origin of the qualitative difference between the two situations with and without the $\eta$ terms,
here we take the bilayer case for illustration. When $N_{z}=2$, in the basis $(c_{a,\uparrow,k_{x},k_{y},z=1}^{\dag}
,c_{b,\uparrow,k_{x},k_{y},z=1}^{\dag},c_{a,\downarrow,k_{x},k_{y},z=1}^{\dag},c_{b,\downarrow,k_{x},k_{y},z=1}^{\dag},\\
c_{a,\uparrow,k_{x},k_{y},z=2}^{\dag}
,c_{b,\uparrow,k_{x},k_{y},z=2}^{\dag},c_{a,\downarrow,k_{x},k_{y},z=2}^{\dag},c_{b,\downarrow,k_{x},k_{y},z=2}^{\dag})$, the normal-state Hamiltonian can be written as
\begin{align}
H(\bm{k})&=(m-t\cos k_{x}-t\cos k_{y})\sigma_{z}-\frac{t_{z}}{2}\rho_{x}\sigma_{z}
\nonumber
\\&\quad
+\lambda (\sin k_{x}s_{z}\sigma_{x}- \sin k_{y}\sigma_{y})
+\eta \sin k_{x}\sin k_{y}\rho_{y}s_{y}\sigma_{x}
\nonumber
\\&\quad
+\frac{\eta}{2} (\cos k_{x}-\cos k_{y})\rho_{y}s_{x}\sigma_{x}
+V_{0}\rho_{z},
\end{align}
where the Pauli matrices $\sigma_{i}$, $s_{i}$, and $\rho_{i}$ act on orbital, spin, and layer degrees of freedom,
respectively. When $\eta=0$, although the physical inversion symmetry (the inversion symmetry operator becomes $\tilde{I}=\rho_{x}\sigma_{z}$
as it should exchange the two layers) is broken, one finds that the Hamiltonian still commutes with the antiunitary operator $is_y \mathcal{K}\sigma_z$ which is a combination of time-reversal and inversion symmetry in orbital space.
The combined symmetry obeys $(is_y \mathcal{K}\sigma_z)^{2}=-1$, and the energy bands thus still obey Kramers' degeneracy at each $\bm{k}$. However, once $\eta\neq0$, the two $\eta$ terms are odd under that combined symmetry, and lead to a splitting of Kramers' degeneracy.
\end{appendix}
\bibliography{ref.bib}
\end{document}